\documentclass[twocolumn,prl]{revtex4}

\usepackage{graphicx}
\usepackage{dcolumn}
\usepackage{bm}
\usepackage{multirow}
\usepackage{color}
\usepackage{amsmath}
\usepackage{gensymb}
\usepackage[colorlinks=true]{hyperref}

\definecolor{mblue}{rgb}{0,0.35,0.75}

\begin{document}
\title{Optical spectroscopy of excited exciton states in MoS$_2$ monolayers\\ in van der Waals heterostructures}

\author{C. Robert$^{1}$, M.A. Semina$^{2}$, F. Cadiz$^{1,3}$, M. Manca$^{1}$, E. Courtade$^{1}$, T. Taniguchi$^{4}$, K. Watanabe$^{4}$, H.~Cai$^{5}$, S. Tongay$^{5}$, B. Lassagne$^{1}$, P. Renucci$^{1}$, T. Amand$^{1}$, X. Marie$^{1}$, M.M. Glazov$^{2}$, B. Urbaszek$^{1}$}

\affiliation{%
$^1$Universit\'e de Toulouse, INSA-CNRS-UPS, LPCNO, 135 Av. Rangueil, 31077 Toulouse, France}
\affiliation{$^2$Ioffe Institute, 194021 St.\,Petersburg, Russia}
\affiliation{$^3$ Physique de la mati\` ere condens\' ee, Ecole Polytechnique, CNRS, Universit\' e  Paris Saclay, 91128 Palaiseau, France}
\affiliation{$^4$National Institute for Materials Science, Tsukuba, Ibaraki 305-0044, Japan}
\affiliation{
$^5$ School for Engineering of Matter, Transport and Energy, Arizona State University, Tempe, AZ 85287, USA}

\begin{abstract}
The optical properties of MoS$_2$ monolayers are dominated by excitons, but for spectrally broad optical transitions in monolayers exfoliated directly onto SiO$_2$ substrates detailed information on excited exciton states is inaccessible.  Encapsulation in hexagonal boron nitride (hBN) allows approaching the homogenous exciton linewidth, but interferences in the van der Waals heterostructures make direct comparison between transitions in optical spectra with different oscillator strength more challenging. 
Here we reveal in reflectivity and in photoluminescence excitation spectroscopy the presence of excited states of the A-exciton in MoS$_2$ monolayers encapsulated in hBN layers of calibrated thickness, allowing to extrapolate an exciton binding energy of $\approx220$~meV. We theoretically reproduce the energy separations and oscillator strengths measured in reflectivity by combining the exciton resonances calculated for a screened two-dimensional Coulomb potential with transfer matrix calculations of the reflectivity for the van der Waals structure. Our analysis shows a very different evolution of the exciton oscillator strength with principal quantum number for the screened Coulomb potential as compared to the ideal two-dimensional hydrogen model.
\end{abstract}


\maketitle
Two-dimensional (2D) crystals of transition metal dichalcogenides such as MX$_2$ (M=Mo, W; X=S, Se, Te) are promising atomically thin semiconductors for applications in electronics and optoelectronics \cite{Novoselov:2016a,Geim:2013a,Mak:2010a, Splendiani:2010a, Wang:2012c,Mak:2016a}. The optical properties of transition metal dichalcogenides (TMD) monolayers (MLs) are governed by very robust excitons \cite{Wang:2017b} with a binding energy of the order of 500 meV \cite{He:2014a,Ugeda:2014a,Chernikov:2014a,Ye:2014a,Qiu:2013a,Ramasubramaniam:2012a,Wang:2015b}. The interplay between inversion symmetry breaking and strong spin-orbit coupling in these MLs  results in unique spin/valley properties \cite{Xiao:2012a,Sallen:2012a,Mak:2012a,Kioseoglou:2012a,Cao:2012a,Jones:2013a,Yang:2015a,Schaibley:2016a}.\\
\indent The first reported optical spectroscopy measurements on ML TMDs were performed on the material MoS$_2$ \cite{Splendiani:2010a,Mak:2010a}, motivated by the high natural abundance of the naturally occurring mineral molybdenite \cite{Dickinson:1923a}. 
But due to the spectrally narrower emission lines of ML MoSe$_2$ and WSe$_2$  \cite{Jones:2013a,Dufferwiel:2015a,Wang:2016b}, low temperature optical spectroscopy studies rapidly switched to other synthetized materials of the TMD family \cite{Liu:2015b,Lundt:2016a,Zhang:2017b,Dufferwiel:2015a,srivastava:2015,stier:2015,Wang:2015d,Aivazian:2015a,Hao:2015a}. Well-defined optical transitions have allowed the observation of excited exciton states, and hence the extrapolation of exciton binding energies, in the optical spectrum of WSe$_2$ \cite{He:2014a,Wang:2015b} and of WS$_2$ MLs \cite{Chernikov:2014a,Zhu:2015b, Chernikov:2015a,Hill:2015a}. Knowledge of the energy of the exciton resonances is also crucial for linear and non-linear optics as in second-harmonic generation and also resonant Raman scattering \cite{Carvalho:2015a,Wang:2015b,Li:2014b,seyler:2015,Soubelet:2016a,chakraborty2013layer,Glazov:2017a}.\\
\begin{figure*}
\includegraphics[width=0.80\textwidth,keepaspectratio=true]{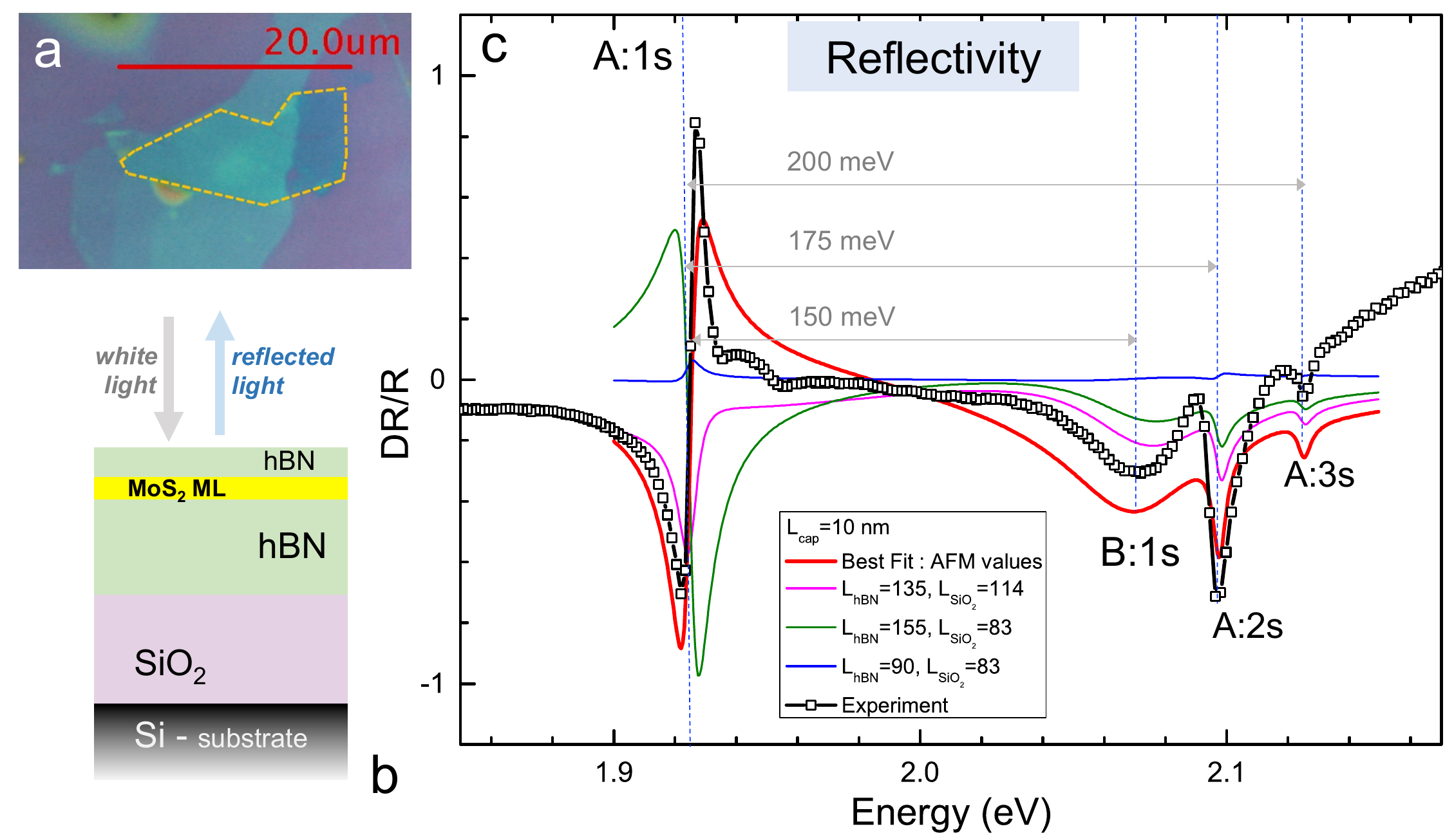}
\caption{\textbf{Optical spectroscopy results.} Sample 1. (a) Optical microscope image of an MoS$_2$ flake encapsulated in hBN. The scale-bar is $20\, \mu$m. (b) Schematic of sample structure, the top (bottom) hBN layers are $7\pm3$~nm ($130\pm5$~nm) thick as determined by AFM, the SiO$_2$ layer is 83~nm thick. (c) Differential reflectivity measurement $(R_{ML}-R_{sub})/R_{sub}$ performed with a power-stabilized white halogen lamp. The strong transition at lower energy is due to the  A-exciton ground state absorption. At higher energies, three more transitions are clearly visible, the broader one is due to the B-exciton ground state  while the other two are tentatively ascribed to be the first two excited states of the A-exciton: A$:2s$ and A$:3s$. Model simulations using the hBN thicknesses as determined by AFM are shown by the bold, red curve. Additional simulations (fine magenta, green and blue) for different hBN and SiO$_2$ thicknesses show how the depth and shape of the exciton resonances depends on the individual layer thicknesses. The exciton resonance parameters are as follows: the energy positions of A:$1s$ exciton was tuned to the observed A:$1s$ peak, energies of excited states are calculated and shown in Fig.~\ref{fig:r3}(a); radiative damping for the ground states $\Gamma_{0,{\rm A:1s}}=\Gamma_{0,{\rm B:1s}}=1$~meV, for excited states found from calculation, Fig.~\ref{fig:r3}(b); non-radiative damping $\Gamma_{\mathrm A}=2.5$~meV, $\Gamma_{\mathrm B} = 25$~meV. Refractive indices: $n_{\rm hBN} =2.2$, $n_{\rm SiO_2} = 1.46$, $n_{\rm Si}=3.5$.}\label{fig:r1} 
\end{figure*}
Optical spectroscopy experiments on excited exciton states plays a crucial role aiming to distinguish between the A-exciton and the B-exciton Rydberg series and other excitonic transitions possibly involving carriers from different valleys in momentum space away from the K-point. For example the $\Gamma$-point in the valence band of MoS$_2$ is situated between the A- and B-valence spin-orbit bands in MoS$_2$~\cite{Kormanyos:2013a,Miwa:2015,Molina:2016a} according to atomistic calculations and angle resolved photo electron spectroscopy.\\ 
\indent Due to the very broad optical transition with a linewidth for MoS$_2$ of up to 50~meV at low temperature when studied without hexagonal BN (hBN)  encapsulation \cite{Korn:2011a,Kioseoglou:2012a,Cao:2012a,Lagarde:2014a,Zeng:2012a,Stier:2016a,Mitioglu:2016a},  information on excited exciton states in ML MoS$_2$ is scarce. \textcite{Hill:2015a} report the presence of excited states in the photoluminescence excitation spectroscopy (PLE) spectrum of ML MoS$_2$ at room temperature. The authors ascribe the observed resonances to the first excited states of the B-exciton, and estimate an exciton binding energy of about $400$~meV for monolayers deposited onto fused silica substrates. The excited states of the A-exciton are predicted to be close in energy to the B-exciton, and therefore are not visible in samples with broad transitions studied so far. \\
\indent Here we present the first measurements of the \textit{excited} exciton states in encapsulated monolayer MoS$_2$ in reflectivity and PLE spectra, using the same encapsulation technique that resulted in ground state exciton transitions of down to 2~meV linewidth \cite{Cadiz:2017a}. In our samples we can spectrally separate optical transitions stemming from the excited A-exciton states from the B-exciton 1s state.  We show in reflectivity measurements and simulations that the thickness of each layer of the van der Waals structure impacts the visibility of the exciton states. We discuss the deviations of the relative oscillator strengths and exciton binding energies of the observed Rydberg series for exciton transitions from the ideal 2D hydrogen problem and extrapolate a exciton binding energy of $\approx220$~meV. Moreover, we examine the possible role of optical transitions away from the K-point of the Brillouin zone. We show efficient valley polarization and coherence initialization for laser energies tuned into resonance with the excited A-exciton states.\\
\begin{figure*}
\includegraphics[width=0.99\textwidth,keepaspectratio=true]{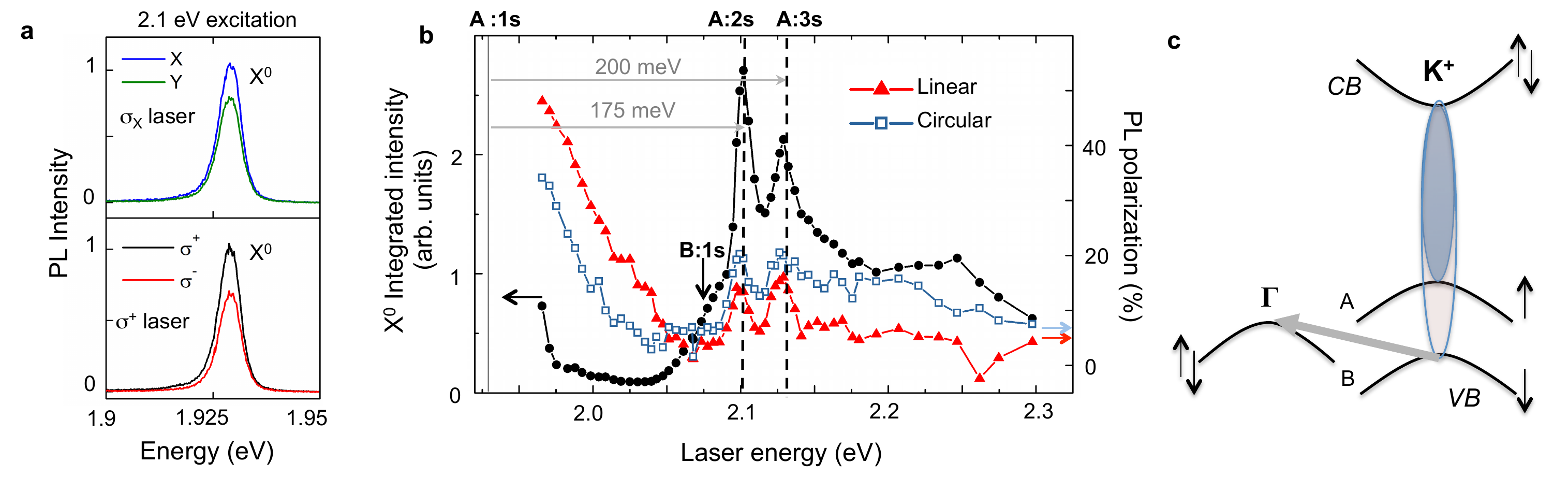}
\caption{Sample 2. (a)  Polarization-resolved photoluminescence at T=4~K following linear (top) and circular (bottom) excitation at $2.1$ eV, exhibiting efficient valley coherence and valley initialization, respectively. (b)PLE measurements. The integrated $X^0$ (i.e. transition A:$1s$) intensity as a function of excitation energy is shown (black circles), as well as the linear (resp. circular) polarization obtained under linear (resp. circular) excitation (red triangles, open squares).  (c)  Schematic single particle band structure of ML MoS$_2$ , for simplicity the small spin splitting in the conduction band (CB) is neglected.}\label{fig:r2} 
\end{figure*}
\indent \textit{Experimental Results.---}
We have investigated MoS$_2$ MLs encapsulated in hBN on top of SiO$_2$/Si substrate, see optical image in Fig.~\ref{fig:r1}(a). In atomic force microscopy (AFM) measurements we determined that the top hBN layer thickness is $7\pm3$~nm, the bottom hBN $130\pm5$~nm thick. The SiO$_2$ layer is 83~nm thick. These van der Waals heterostructures are obtained by mechanical exfoliation of bulk MoS$_2$ (from 2D Semiconductors, USA and growth by chemical vapor transport as in \cite{Cadiz:2017a}) and hBN crystals \cite{Taniguchi:2007a}, following the fabrication technique detailed in Ref. \cite{Cadiz:2017a}.  Encapsulation results in high optical quality samples with well defined optical transitions (FWHM $<5~$meV)  both in photoluminescence (PL) and reflectivity at low temperature, as recently shown \cite{zWang:2017b,Ajayi:2017,Cadiz:2017a, Manca:2017a,Jin:2016a}. 
The PL linewidth of the neutral exciton ($X^0$) thus reaches values down to $2$ meV at cryogenic temperatures, comparable to high quality III-V and II-VI quantum wells grown by molecular beam epitaxy emitting at similar wavelength \cite{Gibbs:2011a,Kreller:1995a,Kasprzak:2006a}. These well-defined emission lines are critical for an in-depth analysis of the optical transitions, since the narrow exciton lines allow us to clearly identify transitions involving the A-exciton excited states as they can be spectrally separated from the B:$1s$ exciton. Here the states are denoted by, e.g., A:$n$s with $n$ being the principal quantum number, in analogy to the $s$-shell (zero angular momentum) states of the hydrogen atom, small mixing of the $s$- and $p$-shell excitons expected in MX$_2$ MLs~\cite{Glazov:2017a,Gong:2017a} is neglected here for simplicity.\\
\indent Figure~\ref{fig:r1}(c) presents the reflectivity spectrum at $T=5$~K of an encapsulated ML, in which we clearly observe the peaks corresponding to the absorption of the lowest energy transition of the A:$1s$ and the B:$1s$ excitons at $1.926$ eV and $2.08$ eV, respectively. The $\sim 150$ meV energy separation between the A and B exciton is in agreement with previous measurements \cite{Luo:2017a,Mak:2012a} and reflects mostly the spin-orbit splitting of the valence band at the K-points of the hexagonal Brillouin zone \cite{Zhu:2011a,Cheiwchanchamnangij:2012a,Miwa:2015}. In addition, higher energy states with measurable oscillator strengths are also visible, that we tentatively ascribe to the two first excited states of the A exciton: A:$2s$ and A:$3s$. \\
\indent These new features above the B:$1s$ transition shown in Fig.~\ref{fig:r1}(c) deserve a detailed analysis. We present now the key results associated to the resonant excitation of the excited exciton states in PL excitation (PLE) with a tunable laser source, see supplement for experimental details. We tune the laser into resonance with the A:$2s$ transition at 2.1~eV. The resulting PL of the A:$1s$ state is strongly co-polarized with the laser as shown in 
Fig.~\ref{fig:r2}(a).  For linearly polarized excitation $\sigma^X$, the resulting A:$1s$ PL is linearly polarized as a consequence of the optical generation of valley coherence i.e. optical alignment of excitons \cite{Jones:2013a,Hao:2015a,Wang:2016b}. 
For circularly polarized laser excitation $\sigma^+$, the PL is co-circularly polarized indicating efficient valley initialization. The fact that a relatively strong polarization is obtained for an excitation laser energy of $\approx175$~meV above the neutral exciton transition is a consequence of a resonance with the A:$2s$ excited state, which relaxes efficiently down to the A:$1s$ state \cite{Wang:2015g}. As shown in Fig.~\ref{fig:r2}(b), the A:$1s$ integrated intensity as a function of excitation energy exhibit two clear and sharp resonances at the A:$2s$ and A:$3s$ energies. Remarkably, the linear (circular) polarization of the A:$1s$ emission under linearly (circularly) polarized excitation also exhibit local maxima when in resonance with the excited A-exciton states. This points towards a fundamental connection between the A:$1s$ and these peaks and is an argument in favor of attributing the two transitions to the A:$2s$ and A:$3s$ states. Both valley coherence and valley polarization reach very high values of $\sim 40\%$ when approaching with the laser the resonance energy of the A:$1s$ transition. Interestingly, the degree of linear polarization is even higher than the circular polarization for excitation energies below $2.05$ eV, as recently observed for nearly resonant excitation \cite{Cadiz:2017a}. This is expected if the exciton spin/valley relaxation process is dominated by Coulomb exchange interaction \cite{Glazov:2014a}.\\
\indent Although the B:$1s$ state is clearly visible in reflectivity, it is much less pronounced in PLE, where B:$1s$ appears as a low energy shoulder of the A:$2s$ transition. A strong signal in PLE relies on efficient absorption at the excitation energy but also on efficient relaxation from this energy to the A:$1s$ state, involving phonon emission \cite{Molina:2011a,Chow:2017b,Jin:2016a}. The weak response in PLE of the A$:1s$ state at the B$:1s$ energy might be due to inefficient relaxation to the A$:1s$ state, with the $\Gamma$ valence states providing an alternative relaxation channel for holes, as sketched in Fig.~\ref{fig:r2}c. This is probably because the energy splitting between the valence $K$ and $\Gamma$ states is only about 100~meV \cite{Kormanyos:2013a,Miwa:2015,Molina:2016a}. Fast B-exciton relaxation could also contribute to the spectral broadening of the transition.\\
\indent To confirm that the transitions we uncover in PLE and reflectivity can be ascribed to neutral and not charged excitons, we have performed experiments on gated structures \cite{Mak:2013a,Ross:2013a}, see supplement, where we also present results on up-conversion and hot PL.\\
\begin{figure*}
\includegraphics[width=0.70\textwidth,keepaspectratio=true]{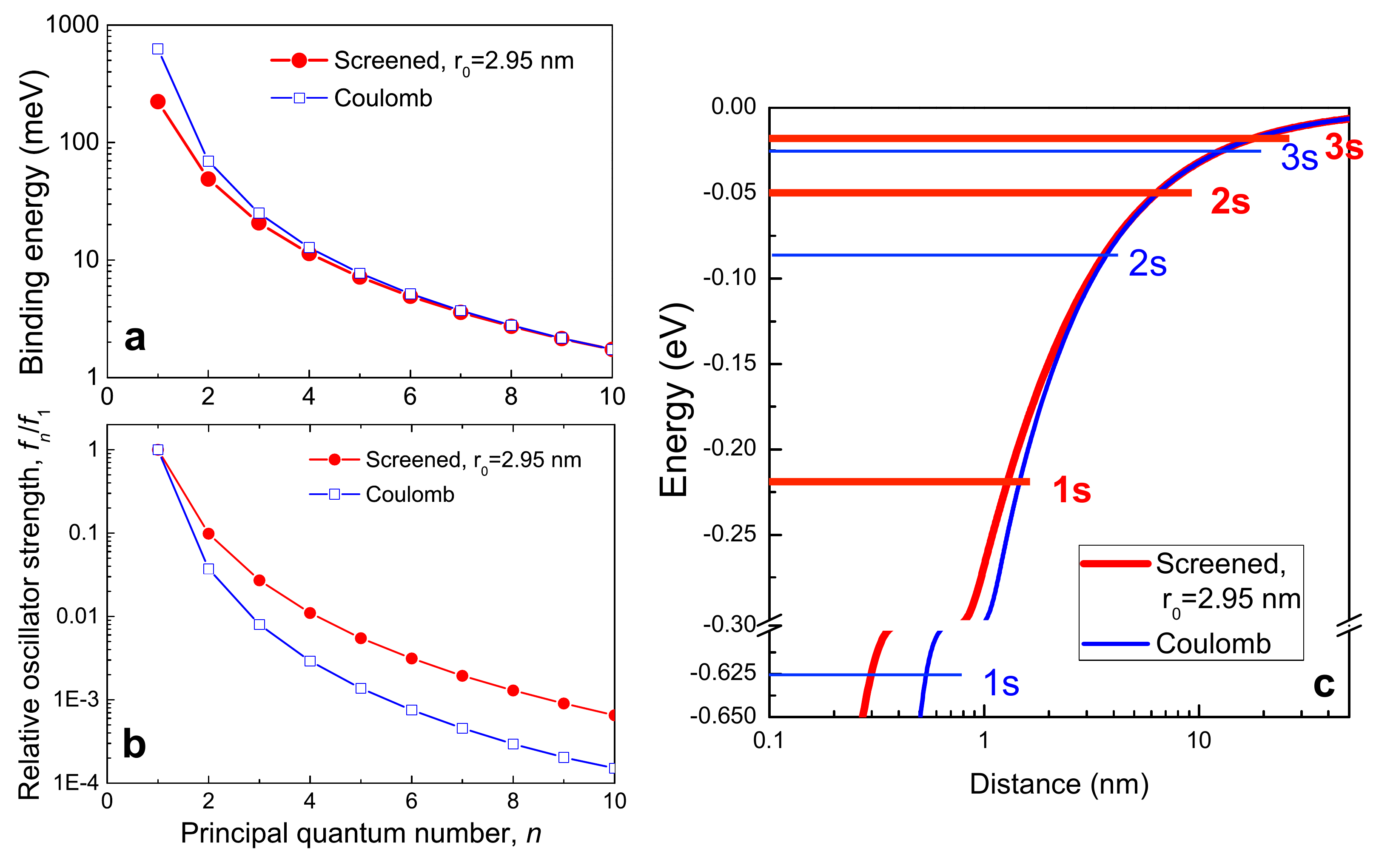}
\caption{\textbf{Results of calculations.} (a) Exciton binding energy for screened $r_0 = 2.95$~nm, Eq.~\eqref{eq1} and unscreened 2D Coulomb potential as a function of principal quantum number $n$ for ML MoSe$_2$. (b) Same as (a), but for the relative oscillator strength normalized at the A:$1s$ exciton oscillator strength. (c) Comparison of the bound states in the screened and unscreened Coulomb potential. }\label{fig:r3} 
\end{figure*}
\indent \textit{Discussion.---}
The fascinating optical properties of TMDC MLs are based mainly on the transitions at the K-point of the Brillouin zone. Here we try to explore optical transitions at higher energy than the exciton ground state. Optical transitions at higher energy can be divided into 3 categories which might spectrally overlap: (1) excited states of the A-exciton, (2) ground state B-exciton, B:$1s$, (3) other transitions in the Brillouin zone, possibly indirect and therefore phonon assisted. The presence of transitions in reflectivity demonstrates non-zero oscillator strength for the absorption, which indicates that phonon-assisted processes are most likely not at the origin of the transitions. Moreover, the PLE data presented above demonstrates a close relation between the higher energy peaks and the ground A:$1s$ exciton states and gives arguments in favor of ascribing the sharp peaks in the reflectivity to the excited states of A-exciton. Below we provide quantitative analysis of the energies and oscillator strengths of the A-exciton Rydberg series and demonstrate its consistency with experimental observations.\\
\indent We performed numerical calculations in order to estimate the exciton binding energy in these encapsulated monolayers. The Coulomb interaction in an encapsulated 2D material  is modelled using the potential \cite{Rytova:1967,Keldysh:1979a,Cudazzo:2011a,PhysRevB.88.045318,Wu:2015a} 
\begin{equation}
V(r)=-\frac{e^2}{8\epsilon_0 r_0} \left[ \mathbf H_0\left( \frac{\kappa r}{r_0} \right) -  Y_0 \left( \frac{\kappa r}{r_0} \right)   \right]
\label{eq1}
\end{equation}
Here $e$ is the electron charge, $\epsilon_0$ is the vacuum permittivity, $\mathbf H_0$ and $Y_0$ are the Struve and Neumann functions, respectively, and $r_0$ is  a screening length characterizing the MoS$_2$ dielectric nature. We then solved the Schr\"odinger equation with the potential in Eq.(\ref{eq1}) by modelling the encapsulation by hBN by an effective relative dielectric constant $\kappa=4.5$ as in ref.\cite{Stier:2017a}, and by using an exciton reduced mass of $\mu=0.25 \;m_0$ \cite{PhysRevB.88.045318}. The results are shown in Fig.~\ref{fig:r3}(a). For a screening length of $r_0=2.95 \pm 0.1$ nm, we obtain the exciton binding energy of $\approx 222$~meV, with the A:$1s$-A:$2s$ separation of $\approx174$~meV and the A:$3s$-A:$2s$  separation of $\approx 28$~meV, in excellent agreement with the values observed in our reflectivity and PLE spectra ($173 \pm 5$ and 28$\pm$ 3~meV, respectively). The screening length is related to the 2D polarizability via $r_0=2\pi \chi_{2D}$, corresponding therefore to $\chi_{2D}=4.47 \pm 0.15$~\AA, in reasonable agreement with theory \cite{PhysRevB.88.045318}.\\ 
\indent Deviations of the electron-hole interaction potential from the $1/r$ law give rise also to the deviations of the oscillator strengths of A:$ns$ states, $f_n$, from the ideal 2D exciton model, where $f_n = f_{1}/(2n-1)^3$ \cite{Shinada:1966a}. This is illustrated in Fig.~\ref{fig:r3} where the ratio $f_n/f_{1}$ is plotted showing strong increase of the relative oscillator strengths for the screened interaction as compared to the standard Coulomb potential. In fact, the ground and first excited states are formed, as shown in Fig.~\ref{fig:r3}(c), in a much shallower, $\propto \log{(r/r_0)}$ effective potential. Hence, the oscillator strength, proportional to the probability to find the electron and the hole within the same unit cell, is smaller for the screened than for the ideal Coulomb potential. 
The higher the principal quantum number $n$ is, the closer is $V(r)$, Eq.~\eqref{eq1}, to its $1/r$ asymtotics. As a result the oscillator strengths approach those in $1/r$ potential. Therefore, the decrease of $f_n/f_1$ is weaker than for the ideal 2D exciton due to smaller oscillator strengths for the low $n$ states. Note that comparatively strong excited exciton state resonances are also predicted from DFT-BSE calculations of the optical response of ML MoS$_2$, as for instance in \cite{Qiu:2013a}. However, in these calculations the exact physical origin of a resonance in absorption is difficult to trace back to a precise quantum number. \\
\indent Although the calculations demonstrate that the relative oscillator strengths for 2s and 3s excitonic states are higher than for the $1/r$ unscreened potential, the amplitude of the A:$2s$ and A:$3s$ resonances in the reflectivity spectra are unexpectedly high and, at a first glance, cannot be accounted for by the increased $f_n/f_1$ in Fig.~\ref{fig:r3}(b). However, the light reflection from our van der Waals sample is determined not only by the monolayer itself, but also by the hBN, SiO$_2$ and Si layers, cf. Ref.~\cite{1742-6596-917-6-062022,Lien:2015}. The cap and bottom layers form a microcavity-like structure enhancing the reflection of the light. To account for the details of light propagation in the sample we employed standard transfer matrix technique, see~\cite{Ivchenko:2005a} and the supplement for details and calculated the reflection contrast spectra for the studied sample. The amplitude reflection coefficient of MoS$_2$ ML was taken in the form
\begin{multline}
r(\hbar\omega) = \sum_{n=1}^3 \frac{\mathrm i \Gamma_{0,\mathrm{A:}n\mathrm{s}}}{E_{\mathrm{A:}n\mathrm{s}} - \hbar\omega - \mathrm i (\Gamma_{0,\mathrm{A:}n\mathrm{s}}+\Gamma_{\rm A})}\\
+ \frac{\mathrm i \Gamma_{0,\mathrm{B:}1\mathrm{s}}}{E_{\mathrm{B:}1\mathrm{s}} - \hbar\omega - \mathrm i (\Gamma_{0,\mathrm{B:}1\mathrm{s}}+\Gamma_{\rm B})}.
\label{eq2}
\end{multline}
It includes independent contributions of the A:$1s$, $2s$ and $3s$ excitons as well as that of the B:$1s$ exciton, $E_{\mathrm{A:}n\mathrm{s}}$, $E_{\mathrm{B:}1\mathrm{s}}$ are the corresponding energies, $\Gamma_0$ and $\Gamma$ are the radiative and non-radiative dampings of the excitons. Equation~\eqref{eq2} assumes that the exciton resonances are well separated.
The interference-enhanced reflection indeed improves the visibility of the excited states and the results of simulation plotted by the red curve in Fig.~\ref{fig:r1} reproduce all features of experimental data rather well. In the calculations the only fitting parameters where absolute energy positions of A:$1s$ and B:$1s$ excitons, its non-radiative dampings and radiative damping of the A:$1s$ exciton. The radiative damping of B:$1s$ exciton was set to be equal to $\Gamma_{0,\text{A:1s}}$.
 All other energies and radiative dampings were found from the calculations presented in Fig.~\ref{fig:r3}. Note that small deviations of the layer thicknesses from the values found in AFM studies (green, blue and magenta curves) result in completely different amplitudes and shapes of features in the reflectivity. This opens the way to control and engineer the optical spectra of the van der Waals heterostructures by choosing appropriate thicknesses of hBN and SiO$_2$ layers resulting in enhancement or suppression of excitonic resonances. \\
\indent As additional identification, magneto-optics in high magnetic fields is desirable to check if the transitions assigned to different Rydberg states have a different diamagnetic shift, as recently demonstrated in magneto-transmission experiments on A:$2s$ to $4s$ states in ML WSe$_2$ \cite{Stier:2017a}. For a quantitative analysis of the oscillator strength of the excited excitons states in the experiment, in addition the impact of mixing of $s-$ and $p-$shell excitonic states needs to be investigated \cite{Glazov:2017a,Gong:2017a}. \\
\indent In conclusion the first direct measurements of the excited states of the A-exciton  in ML MoS$_2$ in reflectivity and PL spectroscopy are reported. These experiments allow to estimate the exciton binding energies and oscillator strengths. The importance of accounting for the light propagation in the multilayer van der Waals heterostructure for quantitative description of the experimental data is demonstrated. \\
\indent \emph{Acknowledgements.---}  
We acknowledge funding from ERC Grant No. 306719., ITN Spin-NANO Marie Sklodowska-Curie grant agreement No 676108, ANR MoS2ValleyControl and ANR VallEx. X.M. also acknowledges the Institut Universitaire de France. K.W. and T.T. acknowledge support from the Elemental Strategy Initiative conducted by the MEXT, Japan and JSPS KAKENHI Grants No. JP26248061, No. JP15K21722, and No. JP25106006. M.A.S. and M.M.G. acknowledge partial support from LIA ILNACS, RFBR projects 17-02-00383, 17-52-16020 and RF President Grant MD-1555.2017.2. S.T acknowledges support from NSF DMR-1552220. 

\newpage
\section{Supplement}
\label{SI}
\subsection{ Experimental Methods}
The top (bottom) hBN layers are $7\pm3$~nm ($130\pm5$~nm) thick as determined by AFM and the in-plane size of the obtained MoS$_2$ ML is typically  $\sim 10\times10$~$\mu$m$^2$. The samples are held on a cold finger in a closed-cycle cryostat.  Attocube X-Y-Z piezo-motors allow for positioning with nm resolution of the ML with respect to the microscope objective (numerical aperture NA$=0.82$) used for excitation and collection of luminescence. For PL measurements, the ML is excited by picosecond pulses (repetition rate $80$ MHz) generated by a tunable frequency-doubled optical parametric oscillator (OPO) synchronously pumped by a mode-locked Ti:Sa laser and focused onto a spot diameter $\approx $ 1 $\mu$m. The PL signal is dispersed in a spectrometer and detected with a Si-CCD cooled camera. The typical excitation power is $5 \; \mu$W. The white light source for reflectivity is a halogen lamp with a stabilized power supply.

\subsection{Additional experimental results}

\begin{figure*}
\includegraphics[width=0.65\textwidth,keepaspectratio=true]{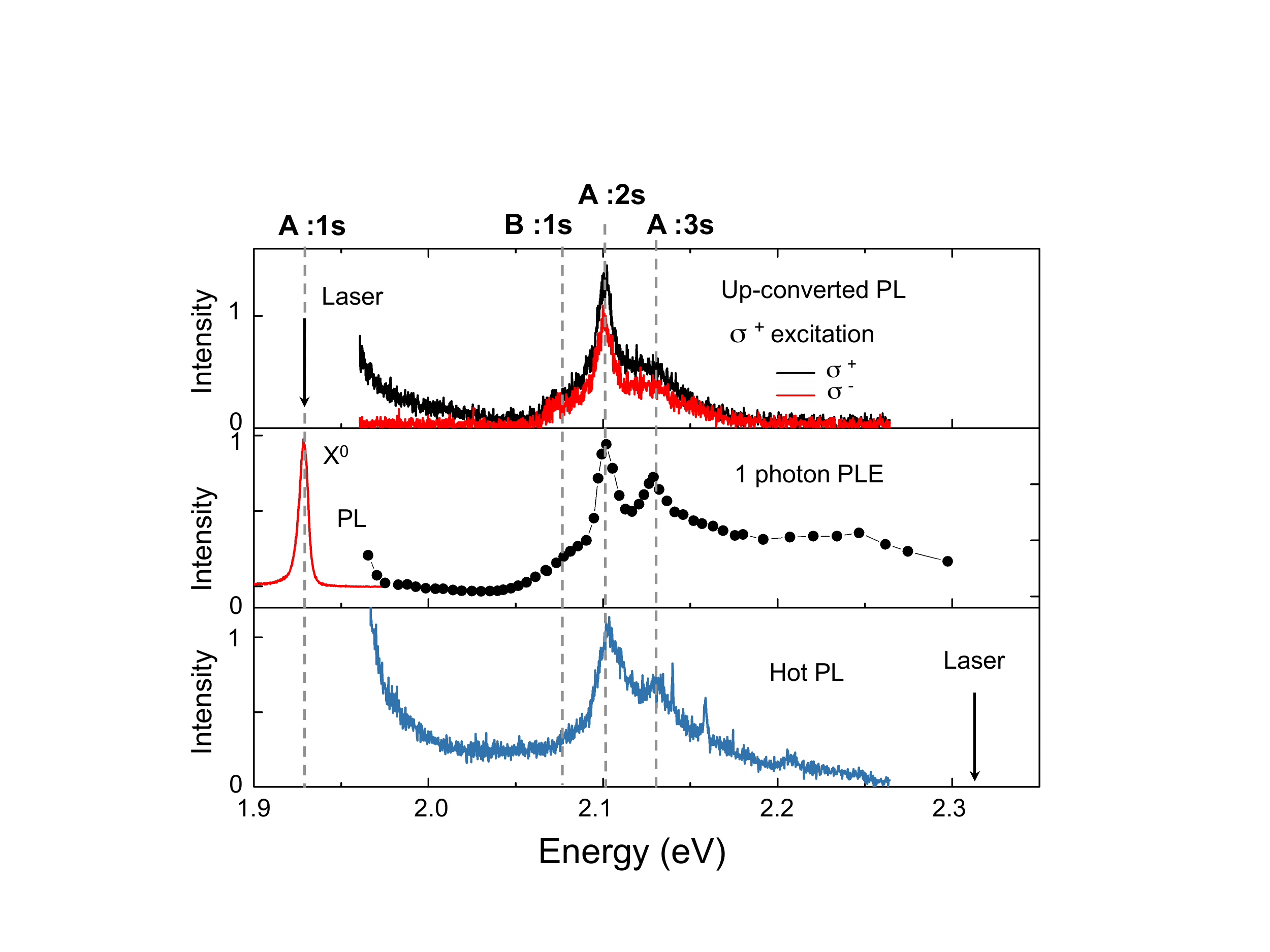}
\caption{\textbf{Exciton upconversion}. $T=4$~K. (a) Up-conversion PL following resonant excitation of the $X^0$ transition with $\sigma^+$ polarized light. (b) $X^0$ PLE spectrum). (c)  Hot PL following a 2.34 eV excitation. }\label{fig:r3} 
\end{figure*}
\begin{figure*}
\includegraphics[width=0.85\textwidth,keepaspectratio=true]{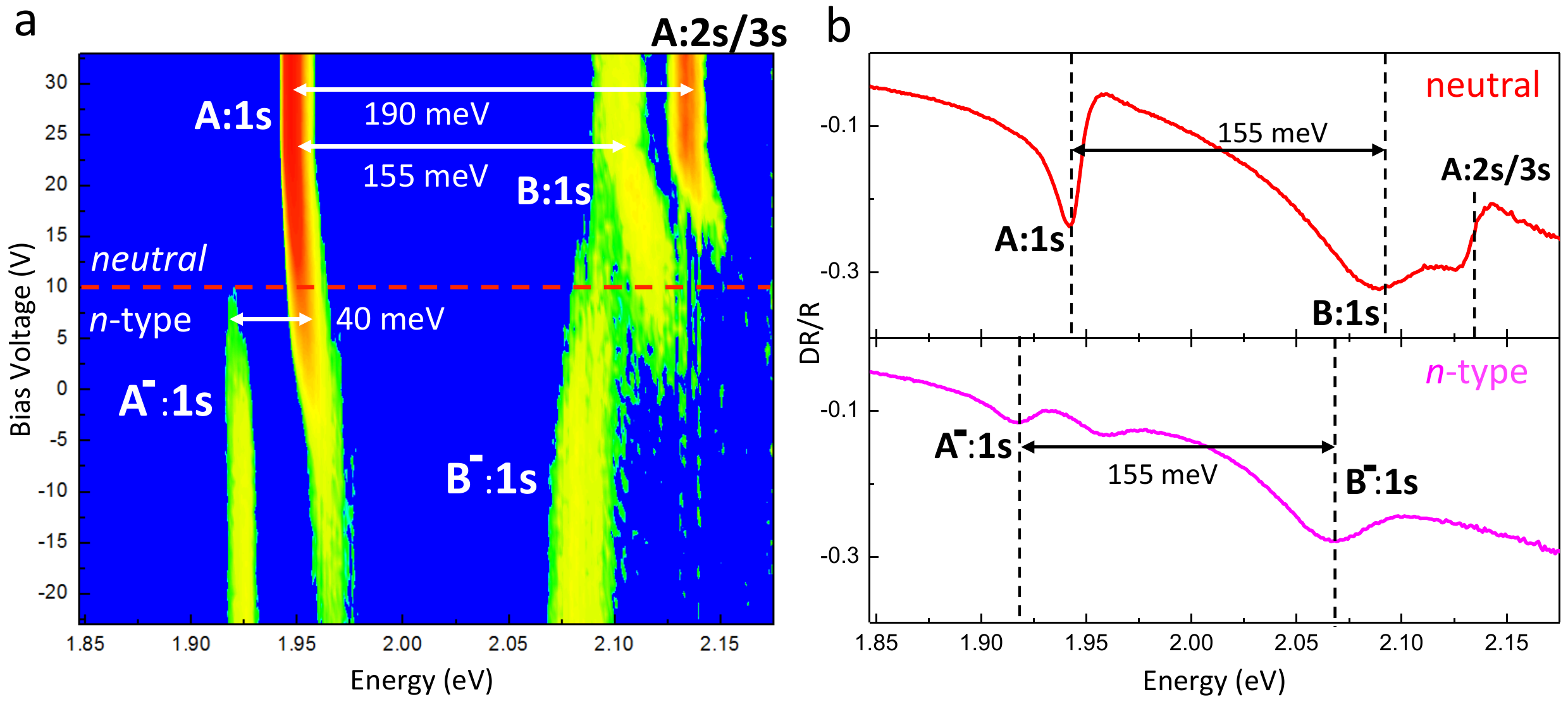}
\caption{\textbf{Charge tunable device}. (a) First derivative of differential reflectivity plotted as a function of gate voltage. The dashed red line indicates the passage from the neutral to the $n$-type regime. The three optical transitions that are clearly visible in the neutral regime decrease in intensity as the $n$-doping is increased, B:1s and the excited states A:2s/3s disappear completely. This suggest that they are associated with neutral exciton states. The two lines appearing in the n-type region are instead identified as negative trions of the A- and B-transitions. (b) Differential reflectivity spectra extracted for the two doping regimes at a bias of +33~V (top) and -23~V (bottom). Note that the energy distance between A:1s and B:1s and the one between A$^-$:1s and B$^-$:1s is the same.}\label{fig:r4} 
\end{figure*}
\indent In Fig. \ref{fig:r3} we present the results of the up-conversion experiment. Excitation of the exciton transition A:1s results surprisingly in PL emission up to 200~meV \textit{above} the laser excitation energy. The up-converted PL emission occurs at the A:2s and A:3s energies, as can be seen by direct comparison with PLE and hot PL presented on the same figure. The exact microscopic origin for the up-conversion is still under investigation, but studies in ML WSe$_2$ \cite{Manca:2017a} hint at scenarios involving resonant two-photon absorption and Auger-type shake up processes. Our results introduce PL up-conversion as an additional technique for studying excited exciton states in ML MoS$_2$. For $\sigma^+$ polarized A:1s excitation we detect a slightly $\sigma^+$ polarized up-conversion emission, indicating that valley initialization is not completely lost in the up-conversion process.  \\
\indent To confirm that the transitions we uncover in PLE, reflectivity, up-conversion and hot PL can be ascribed to neutral and rather than to charged excitons, we have performed experiments on gated structures \cite{Mak:2013a,Ross:2013a}. We present differential reflectivity spectra for a charge-tunable MoS$_2$ ML in Fig.~\ref{fig:r4}. Panel (a) shows a color-plot of the first derivative of differential reflectivity as a function of gate voltage.These results clearly show that the A:2s transition is only visible in the neutral regime for a bias greater that 10~V for this particular sample. Increasing the electron concentration results in an transfer of the oscillator strength from neutral complexes to negatively charged trions, for both A and B excitons. The negatively charged A$^-$ transition appears in energy about 40~meV below the neutral A-exciton. These results demonstrate that the peaks attributed to the $A$ excited states correspond indeed to neutral quasiparticles. \\
\subsection{Transfer matrix method}
In order to calculate the reflection coefficient of the van der Waals heterostructure we use the transfer matrix formalism. We consider the normal incidence of radiation and use the basis of the waves propagating in the positive and negative $z$-directions and present the transfer matrices of individual structure elements as follows:\\
\begin{subequations}
\emph{MoS$_2$ ML:}
\begin{equation}
\hat T_{ML} = \frac{1}{t}\begin{pmatrix}
{t^2-r^2} & r\\
-r & 1
\end{pmatrix},
\end{equation}
where $r$ is the reflection coefficient of the ML, see Eq.~(2) of the main text, $t=1+r$ is the amplitude transmission coefficient of the ML;\\
\emph{homogeneous layer:}
\begin{equation}
\hat T_{ML} = \begin{pmatrix}
\exp{(\mathrm i kL)}  & 0\\
0 & \exp{(-\mathrm i kL)}
\end{pmatrix},
\end{equation}
where $k= \omega n/c$, $n$ is the  refractive index $n$ and $L$ is the layer thickness;\\
\emph{interface between layers:}
\begin{equation}
\hat T_{n_1\to n_2} = \frac{1}{2n_1} \begin{pmatrix}
n_1+n_2 & n_2 - n_1\\
n_2 - n_1 & n_2+n_1
\end{pmatrix},
\end{equation}
the light falls from the layer with the refractive index $n_1$ to the layer with the refractive index $n_2$.
\end{subequations}
The transfer matrix of our heterostructure reads
\begin{multline}
\label{T:tot}
\hat T_{tot} = \hat T_{\rm SiO_2 \to Si} \hat T_{\rm SiO_2} \hat T_{\rm hBN \to SiO_2}  \hat T_{\rm hBN}'  \\
\times T_{air\to \rm hBN} \hat T_{\rm MoS_2} \hat T_{air\to \rm hBN}^{-1} \hat T_{\rm hBN} \hat T_{air\to \rm hBN},
\end{multline}
where prime denotes hBN substrate layer. The inclusion of extra factors $T_{air\to \rm hBN}$ and $\hat T_{air\to \rm hBN}^{-1}$ around $\hat T_{\rm MoS_2}$ is because we refer the excitonic parameters in TMD ML to the layer in the free space (these factors do not substantially change the calculated reflection contrast). We also do not take into account the background dielectric contrast between the ML and the air. The Si layer is assumed to be thicker than the absorption length, hence, the reflection of light at the interface Si and air is disregarded. The transfer matrix provides the following relation between the  $r_{tot}$ and $t_{tot}$ are the amplitude reflection and transmission coefficients through the structure:
\begin{equation}
\label{T:tot1}
\hat T_{tot}
\begin{pmatrix}
1\\
r_{tot}
\end{pmatrix} = \begin{pmatrix}
t_{tot}\\
0
\end{pmatrix}.
\end{equation}
Equation~\eqref{T:tot1} allows us to obtain $r_{tot}$ and $t_{tot}$ from the transfer matrix. Note that the absorbance of the monolayer can be expressed as $\mathcal A = 1- |r_{tot}|^2 - |t_{tot}|^2/n_{\rm Si}$.

\end{document}